%% file: PROCEEDINGS_CSQCD3_CASALI2.tex
\documentclass[12pt]{article}
\usepackage{epsfig}

\input econfmacros-csqcd3.tex
\def\Title#1{\begin{center} {\Large {\bf #1} } \end{center}}

\begin{document}

\Title{Density dependent magnetic field and the equation of state of hyperonic matter}

\bigskip\bigskip


\begin{raggedright}

{\it Rudiney Hoffmann Casali\index{Vader, D.}, \it D\'ebora Peres Menezes\index{Vader, D.}\\
Departamento de F\'isica\\
Universidade Federal de Santa Catarina\\
88040-900 Campus Universit\'ario\\
Florian\'opolis, SC\\
Brazil\\
{\tt Email: rcasali@fisica.ufsc.br}}
\bigskip\bigskip
\end{raggedright}

\section{Introduction}

\begin{figure*}[h]
\begin{center}
\includegraphics[width=0.5\linewidth,angle=-90]{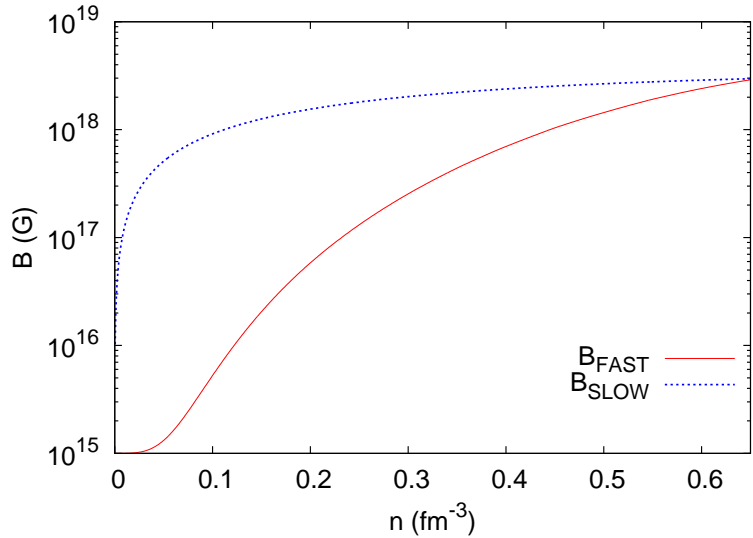}
\caption{Density dependent magnetic field decays, a fast one with $\gamma=4.00$ and $\beta=0.006$, and a slow one with $\gamma=1.00$ and $\beta=0.550$.}
\label{Field-B}
\end{center}
\end{figure*}

We are interested on the effects, caused by strong variable density dependent magnetic fields, on hyperonic matter, its symmetry energy \cite{Matsui},  equations of state
 and mass-radius relations. The inclusion of the anomalous magnetic moment \cite{Broderick}  of the particles involved in a stellar system is performed, and some results are compared with the cases that do not take this correction under consideration. 
The Lagrangian density used follows the nonlinear Walecka model plus the leptons subjected to an external magnetic field as in \cite{Dense_stellar_matter_trapped_neutrinos_under_strong_mf}.
We define total energy density, total pressure, symmetry energy and the magnetic field applied, whose curves can be seen in Figure \ref{Field-B}, as it follows, where  $n_{p}$, $n_{n}$ and $n$ are the proton, neutron and baryonic densities, $n_{0}=0.153~fm^{-3}$ is the saturation density. The model used is the GM1 \cite{Glendening}, $B_{surf}$ is the magnetic field at the surface, taken equal to $10^{15}~G$, $B_{0}$ is the magnetic field for large values of densities. The remaining parameters represent two conditions of magnetic field decay, a fast one with $\gamma=4.00$ and $\beta=0.006$, and a slow one with $\gamma=1.00$ and $\beta=0.550$. The unit of the magnetic field is the critical field $B_{e}^{c}=4.414\times 10^{13}~G$, so that $B=B_{0}/B_{e}^{c}$.
\begin{eqnarray}\label{Eq-energia}
&&\varepsilon_{T}=\varepsilon_{m}+\frac{\left[B\left( \frac{n}{n_{0}} \right) \right]^{2}}{2};P_{T}=P_{m}+\frac{\left[B\left( \frac{n}{n_{0}} \right) \right]^{2}}{2},\\ \
&&a_{sym}=\frac{1}{2}\bigg(\frac{\partial^{2}\varepsilon/n}{\partial t^{2}}\bigg)\Big\vert_{t=0};~t=\frac{n_{n}-n_{p}}{n_{p}+n_{n}},\\ \
&&B\left( \frac{n}{n_{0}} \right)=B_{surf}+B_{0}\left\{ 1-\mathrm{exp}\left[-\beta\left( \frac{n}{n_{0}} \right)^{\gamma} \right] \right\}.
\end{eqnarray}

\begin{figure*}[h]
\begin{center}
\begin{tabular}{ll}
\includegraphics[width=0.33\linewidth,angle=-90]{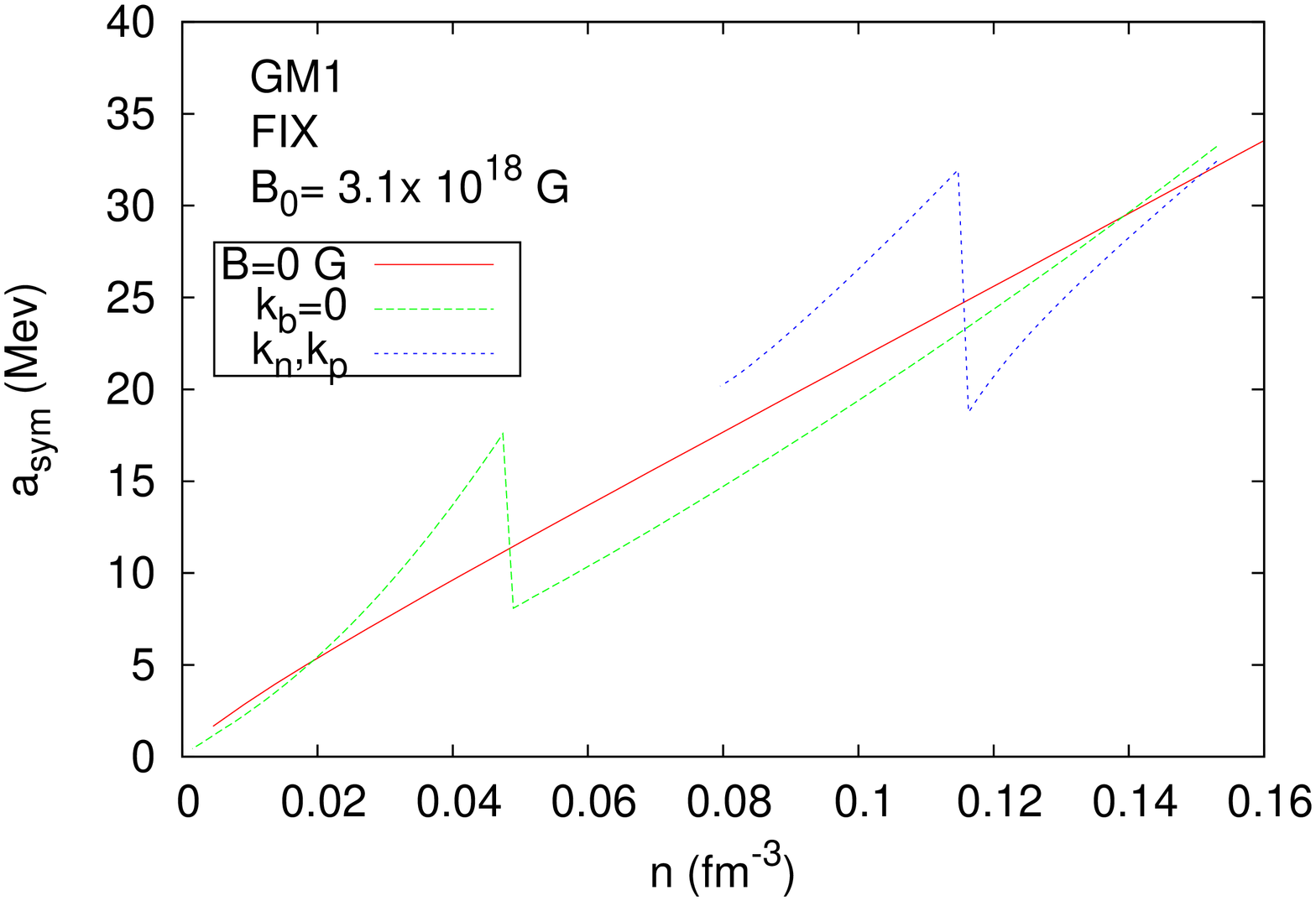}& 
\includegraphics[width=0.33\linewidth,angle=-90]{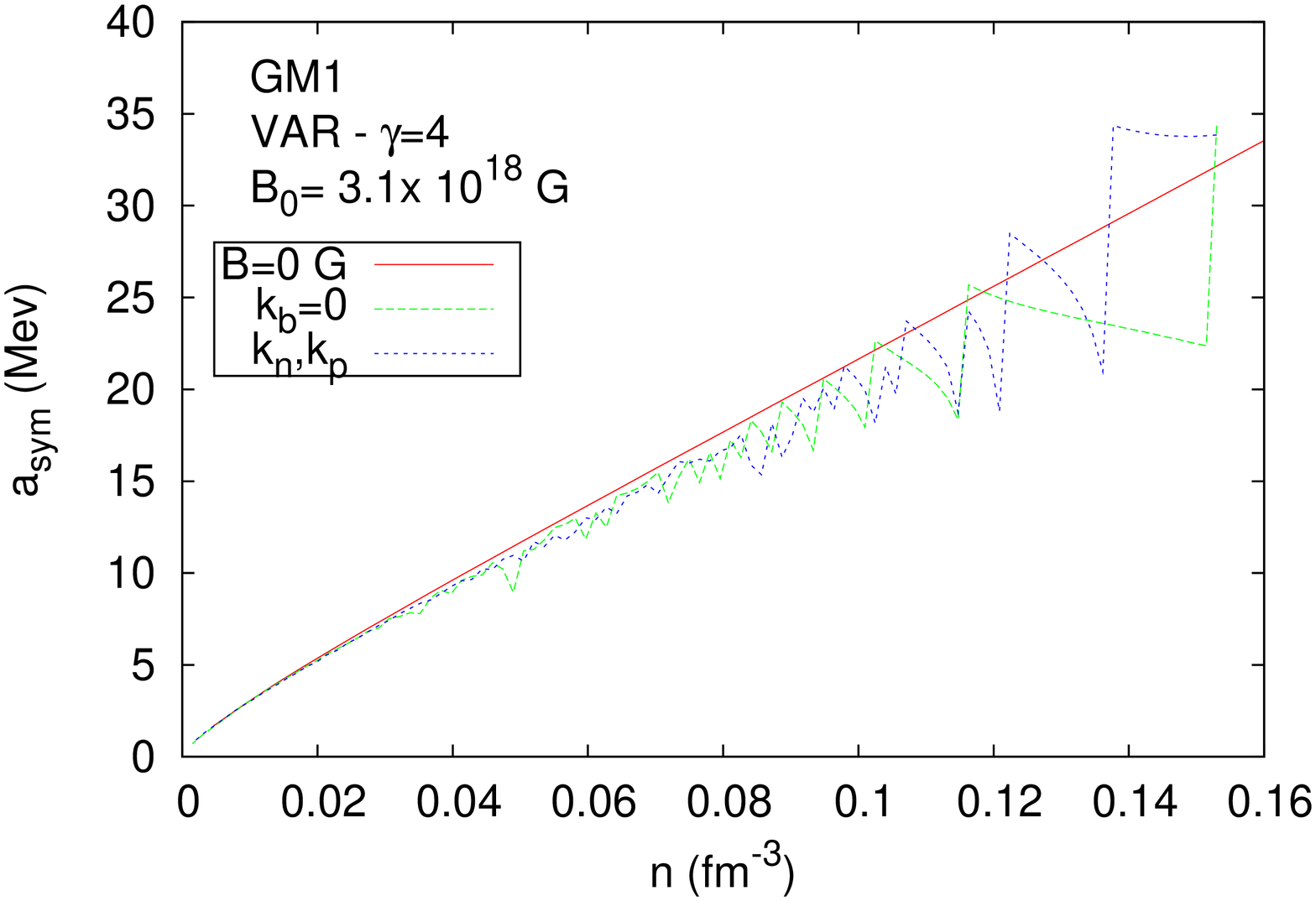}
\end{tabular}
\caption{Symmetry energy for nucleonic matter. Two cases of anomalous magnetic moments are considered, for fixed (left panel) and variable density dependent (right panel) magnetic field.}
\label{Sy-En}
\end{center}
\end{figure*}

\begin{figure*}[h]
\begin{center}
\begin{tabular}{ll}
\includegraphics[width=0.33\linewidth,angle=-90]{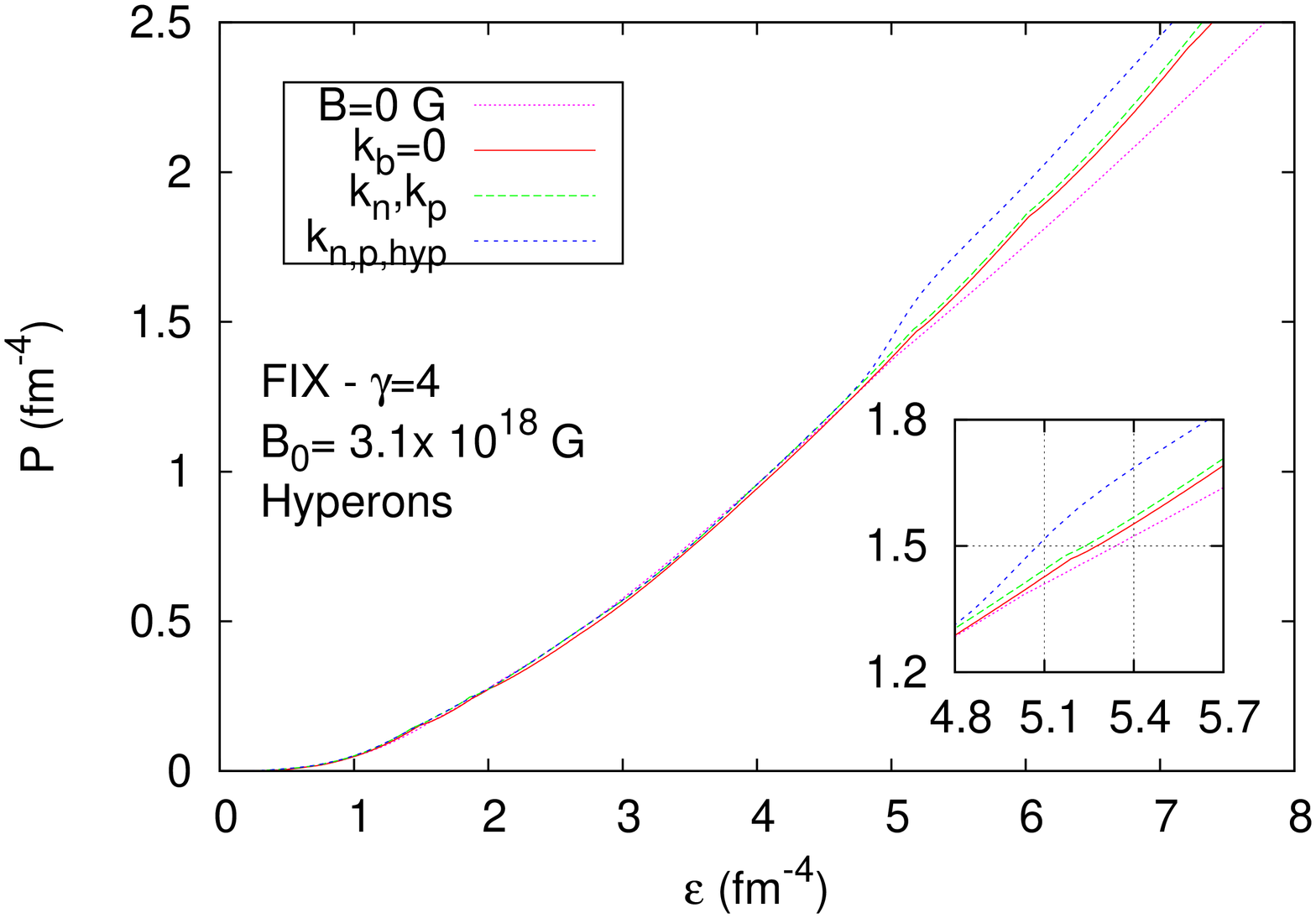}& 
\includegraphics[width=0.33\linewidth,angle=-90]{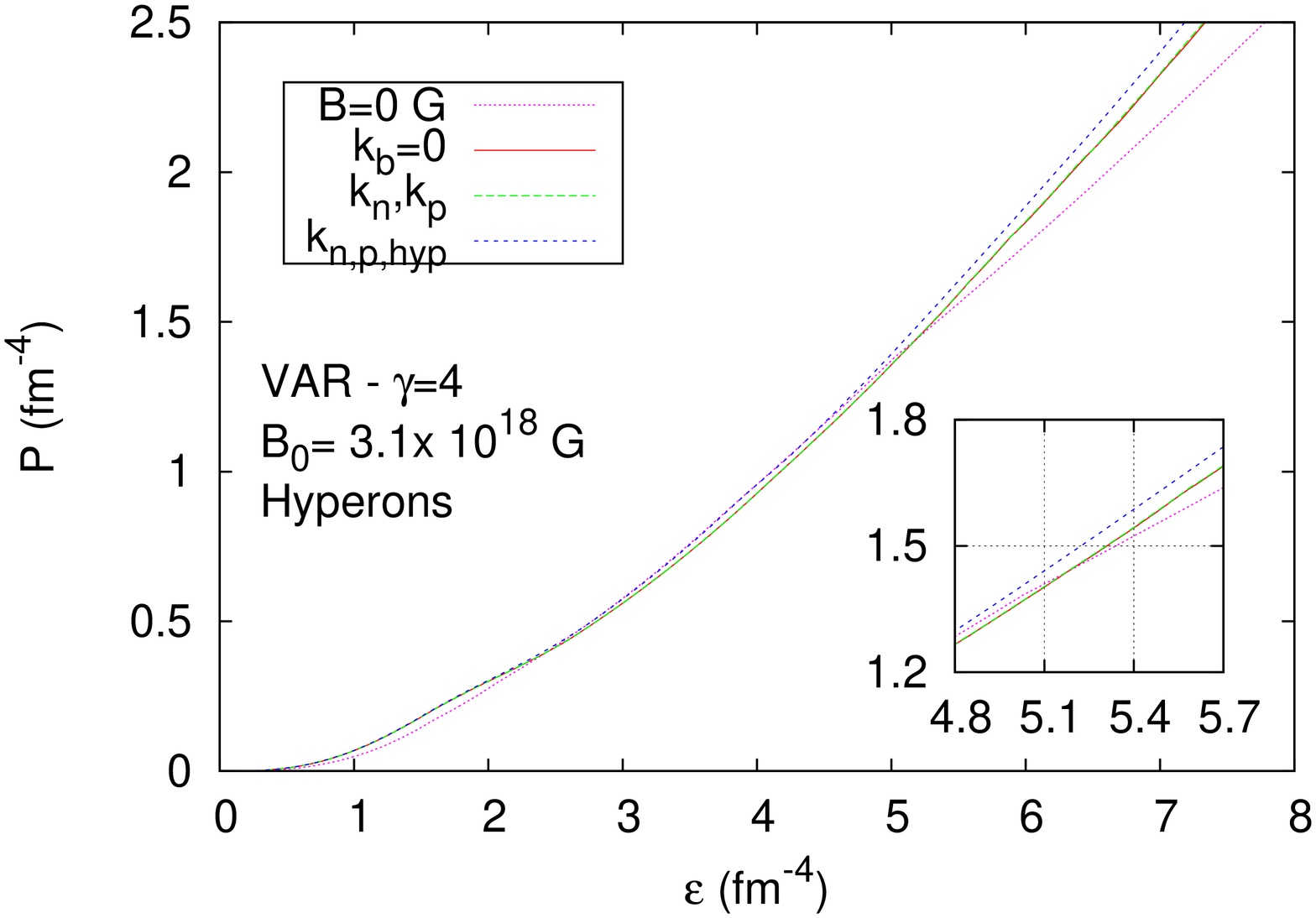}
\end{tabular}
\caption{Equations of state for hyperonic matter. Three cases of anomalous magnetic moment inclusion are considered, for fixed (left panel) and variable density dependent (right panel) magnetic field.}
\label{EOS}
\end{center}
\end{figure*}

\begin{figure*}[h]
\begin{center}
\begin{tabular}{ll}
\includegraphics[width=0.33\linewidth,angle=-90]{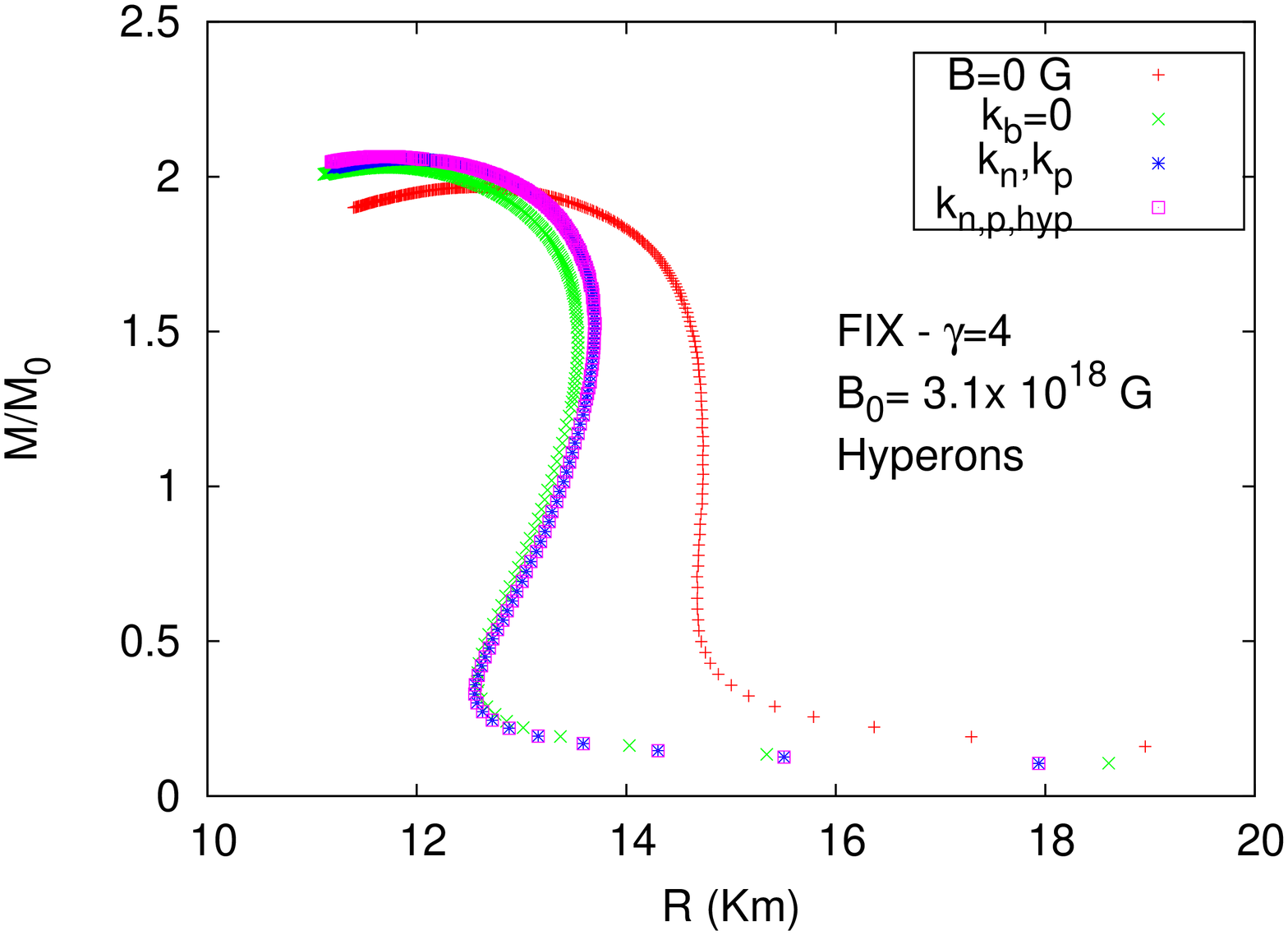}& 
\includegraphics[width=0.33\linewidth,angle=-90]{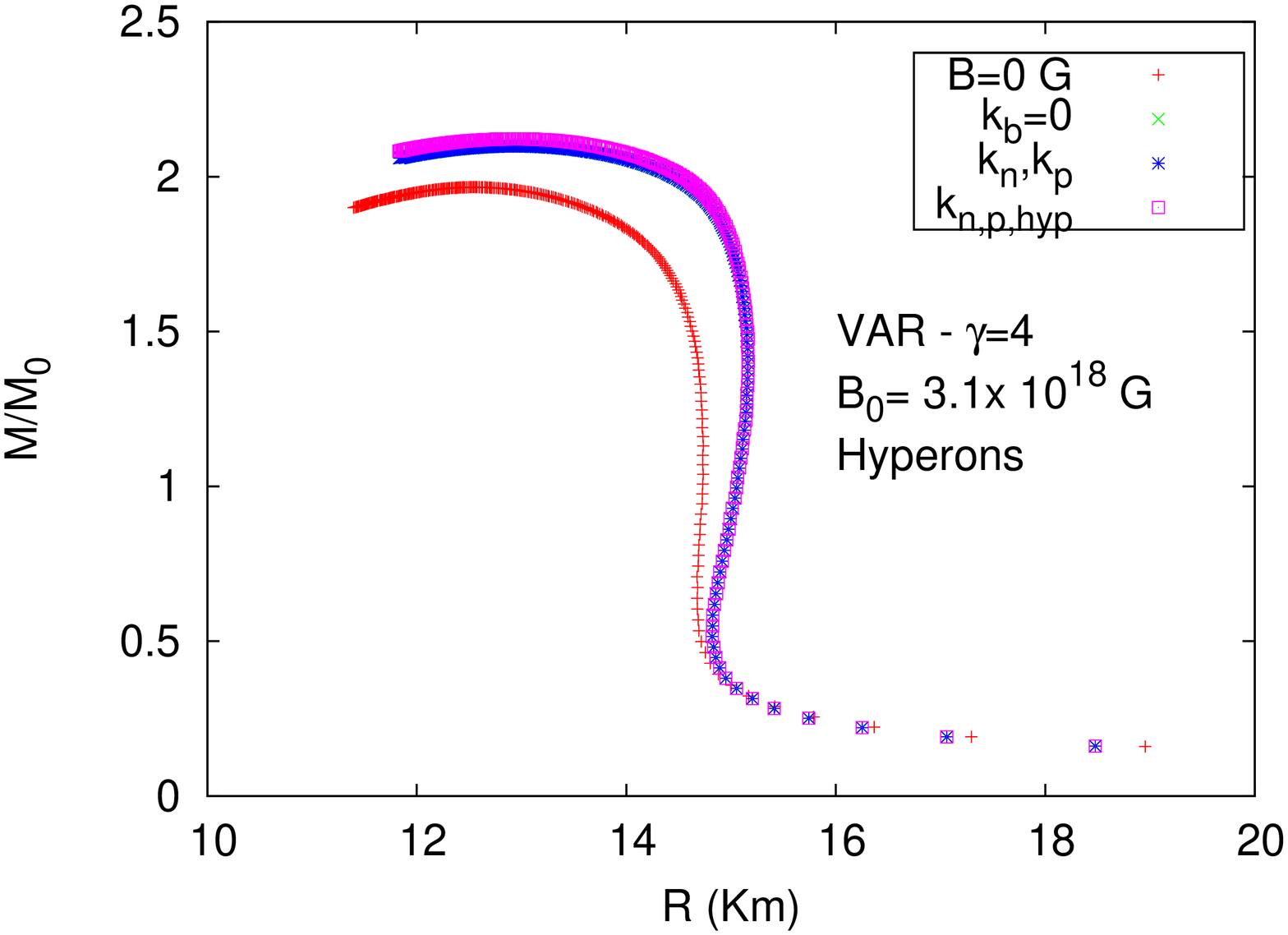}
\end{tabular}
\caption{Mass-radius relation for hyperonic matter. Three cases of anomalous magnetic moments inclusion are considered, for fixed (left panel) and variable density dependent(right panel) magnetic field.}
\label{TOV}
\end{center}
\end{figure*}

In this work we study the behavior of baryonic matter under two types of magnetic field influence: in one of them we consider a fixed magnetic field $B_{0}$ in the equations of state (FIX). In the other one, a variable density dependent magnetic field $B\left( \frac{n}{n_{0}} \right)$ is considered on the equations of state (VAR). Both cases consider a squared density dependent magnetic field term $\left[B\left( \frac{n}{n_{0}} \right) \right]^{2}$ term on the total energy density and total pressure equations. The first two graphics, Figure \ref{Sy-En}, show how the symmetry energy is affected by the magnetic field. On the left a fixed field $B$, on the right a density dependent one. As the magnetic field decays, on the right, it is possible to see the diminishing effects of the Landau level quantization, toward lower densities. Figure \ref{EOS} presents the equation of state of a hyperonic matter. Both cases present smoothening tendencies, as the $\left[B\left( \frac{n}{n_{0}} \right) \right]^{2}$ term decays, but due to the inclusion of the variable density dependent magnetic field also on the equations of state, in the right panel, this tendency is more strong. The effects of the anomalous magnetic moment corrections can be seen in the zoomed areas. 
At the end, in Figure \ref{TOV}, we plot the mass-radius relations, showing results not very different from those of current literature.

\section{Conclusions}

In this work we studied the effects of the inclusion of a density dependent magnetic field in the equations of state and we compared it with the one found in the literature. We notice that the variable density dependent magnetic field generates a much less bizarre energy symmetry curve, as the magnetic field decreases. 
We performed three scenarios for the inclusion of the anomalous magnetic moment corrections, calculated its equations of state and mass-radius relations. It can be seen in Table \ref{table4} that  a variable density dependent magnetic field can be applied on the equations of state without causing great discrepancies from that found in literature. 

\begin{table}[h]
 \centering
\begin{tabular}{|c|c|c|c|}
  \hline
Magnetic Field &AMM&$M_{Max}(M_{0})$&R (Km)\\ \hline
$B=0~G$              &                              & 1.97 & 12.55 \\ \hline
                               & $\kappa_{b}=0$& 2.03& 11.72 \\ \cline{2-4}
  $B_{0}=3.1\times 10^{18}~G$ - FIX& $\kappa_{n,p}$& 2.06& 11.91 \\ \cline{2-4}
                               & $\kappa_{n,p,hyp}$& 2.06& 11.69 \\ \cline{1-4}
  \hline
                               & $\kappa_{b}=0$& 2.10& 12.94 \\ \cline{2-4}
  $B_{0}=3.1\times 10^{18}~G$ - VAR& $\kappa_{n,p}$& 2.10& 12.94 \\ \cline{2-4}
                               & $\kappa_{n,p,hyp}$& 2.12& 12.94 \\ \cline{1-4}
  \hline
 \end{tabular}
 \caption{\label{table4} Mass-radius curves for hyperonic matter summarized. Curves with no corrections on the anomalous magnetic
moment ($k_{b}=0$), with corrections on neutrons and protons ($k_{n,p}$) and with corrections on neutrons, protons and hyperons ($k_{n,p,hyp}$).}\label{table4}

\end{table}

This work was supported by CAPES/CNPq/FAPESC and CAPES/FCT 232/09.

\end{document}

%% file: econfmacros-csqcd3.tex
\def\beq{\begin{equation}}
\def\eeq#1{\label{#1}\end{equation}}
\def\eeqn{\end{equation}}

\def\beqa{\begin{eqnarray}}
\def\eeqa#1{\label{#1}\end{eqnarray}}
\def\eeqan{\end{eqnarray}}

\let\bar=\overbar

\def\Dslash{\not{\hbox{\kern-4pt $D$}}}
\def\dslash{\not{\hbox{\kern-2pt $\del$}}}

\def\msb{{\bar{\ssstyle M \kern -1pt S}}}

\usepackage{fancyhdr,graphicx}